\documentclass[11pt,a4paper]{article}
\usepackage{epsfig}

\newcommand{\alps}	{\ensuremath{\alpha_S}}

\newcommand{\df}	{\ensuremath{\mathrm{d}}}
\newcommand{\e}		{\ensuremath{\mathrm{e}}}

\newcommand{\vc}	{\ensuremath{\mathbf{c}}}
\newcommand{\vb}	{\ensuremath{\mathbf{b}}}
\newcommand{\vl}	{\ensuremath{\mathbf{\ell}}}
\newcommand{\vq}	{\ensuremath{\mathbf{q}}}
\newcommand{\vr}	{\ensuremath{\mathbf{r}}}
\newcommand{\vrc}	{\ensuremath{\mathbf{R}}}

\newcommand{\ap}	{\ensuremath{\alpha_{\mathcal{P}}}}
\newcommand{\apm}	{\ensuremath{(\ap-1)}}

\begin{document}

\begin{titlepage}
\renewcommand{\thefootnote}{\fnsymbol{footnote}}
\begin{flushright}
Cavendish-HEP-95/05 \\
hep-ph/9509353 \\
September 1995
\end{flushright}
\vspace*{20mm}

\begin{center}
\textbf{\Large Studies of Unitarity at Small~$x$ Using
the Dipole Formulation\footnote{Research supported by the UK Particle
Physics and Astronomy Research Council}}
\vspace{10mm}

\textbf{G.P.~Salam} \\
\vspace{3mm}

\textit{Cavendish Laboratory, Cambridge University,} \\
\textit{Madingley Road, Cambridge CB3 0HE, UK} \\
\vspace{2mm}
e-mail: salam@hep.phy.cam.ac.uk
\end{center}

\vspace{20mm}
\begin{abstract}
\noindent Mueller's dipole formulation of onium-onium scattering is
used to study unitarity corrections to the BFKL power growth at high
energies.  After a short discussion of the spatial distribution of
colour dipoles in a heavy quarkonium and the associated fluctuations,
results are presented showing that the one and two-pomeron
contributions to the total cross section are the same at a rapidity $Y
\simeq 14$. Above this rapidity the large fluctuations in the onium
wave function cause the multiple pomeron series to diverge. Resumming
the series allows one to show that unitarity corrections set in
gradually for the total cross section, which is dominated by rare,
large, configurations of the onia. The elastic cross section comes
mostly from much smaller impact parameters and has significant
unitarity corrections starting at a rapidity $Y\simeq 8$.
\end{abstract}

\end{titlepage}

\section{Introduction}
The BFKL pomeron \cite{BaLi78,KuLF77,Lipa86} causes the amplitudes for
certain high energy processes, which are also hard processes, to rise
as a power of the centre of mass energy. At very high energies this
rapid rise will cause the amplitudes to violate unitarity. Unitarity
is the statement that for any given initial states and impact
parameter, the probability of interaction cannot be larger than one.
Because it is a limit on the interaction for each impact parameter,
one cannot tell when an amplitude violates unitarity without knowing
how the amplitude is distributed in impact parameter. This has also
been one of the main uncertainties in calculations (for example
\cite{KwMS91}) of the shadowing corrections to BFKL evolution using
the GLR equation \cite{GrLR83}. However knowledge of the average
interaction for a given impact parameter is insufficient. The wave
functions for each of the colliding objects consist of a variety of
states each with some amplitude. What can happen is that for a given
impact parameter the average interaction can be dominated by
occasional rare combinations of states which have a large interaction
and therefore large unitarity corrections. However because these
configurations are rare, the average interaction can still be small
and one can be misled into thinking that there should be no unitarity
corrections.

One of the simplest frameworks in which to study the BFKL pomeron, and
the onset of unitarity corrections, is the dipole formulation,
proposed by Mueller \cite{Muel94a, MuPa94, Muel94b, Muel95}. For
sufficiently heavy onia, $\alps$, the strong coupling constant, is
small and one can apply perturbation theory. One starts with a
$q\bar{q}$ state and considers the correction to the wave function
which comes from adding one gluon, causing the initial $q\bar{q}$
colour dipole to branch into two colour dipoles ($qg$ and
$g\bar{q}$). In the small-$x$, leading $N_C$ approximation, each of
these dipoles can then branch again, independently, and the process
repeats itself, building up a cascade of colour dipoles. One can then
calculate the onium-onium elastic amplitude from the interactions
between the dipoles of the two onia. The lowest order interaction is
the exchange of a colour neutral pair of gluons. This is one pomeron
exchange. Exchange of larger numbers of pairs of gluons corresponds to
multiple pomeron exchange. This approach is particularly suited to
the study of unitarity because it is formulated in impact parameter
and also because it contains all the information on the fluctuations
in the onium configurations. Further, because the onium wave functions
are each evolved to only half the total rapidity, unitarity
corrections to the cross section should set in long before the density
of partons rises to the point where corrections to the wave functions
need also be taken into account, i.e.\ long before the wave functions
saturate.

In this paper, a Monte Carlo simulation will be used to calculate the
full, unitarised, amplitude for high energy onium-onium elastic
scattering.  Section~\ref{sc:intonon} gives a brief review of the
dipole formalism. Section~\ref{sc:wvfn} discusses certain properties
of the onium wave function, showing that the distribution of dipoles is
more central than had been suggested in \cite{Muel94b}, and
giving more information on the exponential tails of the fluctuations
in dipole density, which were first mentioned in \cite{Sala95}. These
fluctuations have as a consequence that the $k$-pomeron series is
divergent, though it can be resummed by interchanging the order of the
summations over $k$ and over configurations of the onia
\cite{Muel94b}. This and other consequences of fluctuations are
illustrated in section~\ref{sc:mcres}, where the main results of the
paper are presented: the $k$-pomeron and unitarised amplitudes for
onium-onium scattering, as a function of rapidity. One and two pomeron
contributions to the total cross section are seen to be equal at a
rapidity of 14, but the unitarised total cross section carries on
growing well beyond this point. The elastic cross section has much
stronger unitarity corrections and the behaviour of the power of the
growth of the unitarised and 1-pomeron elastic cross sections differ
significantly above $Y=8$.

\section{Onium-onium scattering}
\label{sc:intonon}
\subsection{Single pomeron exchange}
The amplitude for exchange of a single pomeron between two onia moving
fast in opposite directions, with a centre-of-mass energy $\sqrt{s}$
is:

\begin{equation}
A(\vr, Y) = -i \int \df^2\vb \df^2\vb' \int \df z \df z'
                \Phi(\vb, z) \Phi(\vb', z') F(\vb, \vb', \vr, Y)
\label{eq:afromf}
\end{equation}

\noindent The rapidity $Y$ is approximately $\log s/M^2$, with $M$
being the onium mass. The amplitude is determined for a fixed relative
impact parameter $\vr$ between the onia and $\Phi(\vb,z)$ is the
square of the heavy quark-antiquark part of the onium wave functions
with $\vb$ the transverse separation between the quark and antiquark
and $z$ the longitudinal momentum fraction of the antiquark. For
single pomeron exchange, $F = F^{(1)}$, with

\begin{equation}
F^{(1)}(\vb, \vb', \vr, Y) =
        - \int \frac {\df^2 \vc}{2\pi c^2} \frac{\df^2 \vc'}{2\pi c'^2}
        \df^2 \vrc \df^2 \vrc'
        f(\vrc - \vrc', \vc, \vc')
        n(c, b, R, y) n(c', b', |\vrc'-\vr|, Y - y)
\label{eq:f1mine}
\end{equation}

\noindent The factors  $n(c, b, R, y)$ are the density of dipoles
of size $c$ at a distance $R$ from a parent onium of size $b$ after
evolution through a rapidity $y$. The terms $1/2\pi c^2$ are just to
compensate for the normalisation of each wave function (which from now
on refers to the gluon distribution of an onium of fixed size). Note
that for one pomeron exchange the division of rapidity between the two
onia (the choice of $y$) is arbitrary. The amplitude (divided by
$(-i)$) for interaction between a dipole of size $\vc$ and one of size
$\vc'$ whose centres are separated by $\vr$, $f(\vr, \vc, \vc')$, is
the following:

\begin{eqnarray}
\int \df^2 \vr \e^{i\vq . \vr} f(\vr, \vc, \vc') =
        \frac{\alps^2}{2}
        \int \frac{\df^2 \vl}{|\vl|^2|\vq -\vl|^2}
\left[\e^{i\vl . \vc/2} - \e^{-i\vl . \vc/2}\right]
\left[\e^{-i\vl . \vc'/2} - \e^{i\vl . \vc'/2}\right]  \nonumber \\
\left[\e^{i(\vq - \vl) . \vc/2} - \e^{-i(\vq - \vl) . \vc/2}\right]
\left[\e^{-i(\vq - \vl) . \vc'/2} - \e^{i(\vq - \vl) . \vc'/2}\right]
\label{eq:frft}
\end{eqnarray}

\noindent This is considered in more detail in the next section. The
normalisation is chosen such that the total cross section is:

\begin{equation}
\sigma_{tot}(Y) = 2 \int \df^2 \vr\; \mathrm{Im} A(\vr, Y)
\end{equation}

\noindent while for the elastic cross section

\begin{equation}
\frac{\df \sigma_{el}}{\df t} = \frac{1}{4\pi} |A(q,Y)|^2,
\label{eq:elvt}
\end{equation}

\noindent with $t = -q^2$ and

\begin{equation}
A(q,Y) = \int \df^2 \vr \e^{i\vq . \vr} A(\vr, Y).
\end{equation}

\noindent The term total amplitude will be used to refer to $F(b, b',
r, Y)$ integrated over $\vr$, which is equal to half the total cross
section (ignoring the integral of eq.~(\ref{eq:afromf})).

In the limit $r \gg b, b'$, and with $\log r^2/bb' \ll kY$, one has
the following expression for $F^{(1)}$

\begin{equation}
F^{(1)}(b, b', r, Y) = - \frac {8\pi \alps^2 b b'}{r^2}
        \log \left(\frac{16r^2}{bb'}\right)
        \frac{\exp[\apm Y - \log^2(16r^2/bb') / kY]}
             {(\pi k y)^{3/2}},
\label{eq:f1sptl}
\end{equation}

\noindent where $\apm = 4\log 2 \alps N_C / \pi$ and $k =
14\alps N_C\zeta(3)/\pi$, $\zeta(n)$ being the Riemann zeta function.

It is convenient to work in impact parameter and with this
normalisation, because the unitarity condition is then simply $|F(\vr,
Y)| \le 1$. It is the exponential dependence on $Y$ which causes this
condition to be violated as $Y$ increases beyond a certain limit.

\subsection{Dipole-Dipole interaction}
\label{sc:dipdip}
In \cite{Muel94b} the dipole-dipole interaction, $f(r,\vc,
\vc')$ was approximated by:

\begin{equation}
f(\vr, \vc, \vc') = \delta^2(\vr)
\pi \alps^2 c_<^2 \left( 1 + \log\frac{c_>}{c_<} \right)
\label{eq:ddint}
\end{equation}

\noindent This is correct if one averages over angles
and integrates over $\vr$, and is therefore perfectly adequate for
studying single pomeron exchange. However to study multiple pomeron
exchange one needs to know the full expression. It is obtained by
performing the inverse Fourier transform in eq.~(\ref{eq:frft}),
noting that $f(q)$, the Fourier transform of the potential, is a
convolution of two identical terms, each of which gives the single
gluon potential between a pair of dipoles. The result is that $f(\vr)$
is then simply the square of the two-dimensional dipole-dipole
potential:

\begin{equation}
f(\vr, \vc, \vc') = \frac{\alps^2}{2} \left[ \log \frac{
         |\vr + \vc/2 - \vc'/2| |\vr - \vc/2 + \vc'/2| }
        {|\vr + \vc/2 + \vc'/2| |\vr - \vc/2 - \vc'/2| }
        \right]^2,
\label{eq:dipdip}
\end{equation}

\noindent as would be expected for the exchange of
two gluons between a pair of dipoles in two dimensions.

\subsection{Multiple pomeron exchange}
To recover the unitarity condition it is necessary to include higher
order terms. The ones which will be considered here are $k$-pomeron
exchange. This corresponds to the exchange of $k$ colour neutral pairs
of gluons between the two onia. As is discussed in \cite{Muel94b} and
reviewed here in section \ref{sc:rapdiv}, at rapidities where
unitarity effects are setting in, it should be safe to neglect
corrections to the wave functions which would arise due to saturation
of the wave functions. To evaluate multiple pomeron exchange, it is
no longer adequate to use the average dipole distribution: it is
necessary to know the higher moments of the wave functions

In \cite{Muel94b}, the equations governing the evolution of the
wave function are expressed using an operator formalism: the operator
$a^{\dagger}(\vr,\vb)$ ($d^{\dagger}(\vr, \vb)$) creates
a dipole at position $\vr$ and with size $\vb$, in the right (left)
moving onium. The basic vertex of the dipole picture is then $V_1$:

\begin{equation}
V_1 = \frac{\alps N_C}{2\pi^2}
\int \frac{b_{01}^2}{b_{02}^2 b_{12}^2}
 \df^2\vb_{01} \df^2\vb_2 \df^2\vr
a^\dagger (\vr + \frac{\vb_{12}}{2}, \vb_{02})
a^\dagger (\vr + \frac{\vb_{02}}{2}, \vb_{12})
a(\vr, \vb_{01}).
\label{eq:v1}
\end{equation}

\noindent This expresses the branching of a colour dipole of size
$\vb_{01}$ at position $\vr$ into the two dipoles which are formed by
the production of a gluon at $\vb_2$. Together with this there are
virtual corrections

\begin{equation}
V_2 = - \frac{\alps N_C}{2\pi^2}
\int \frac{b_{01}^2}{b_{02}^2 b_{12}^2}
 \df^2\vb_{01} \df^2\vb_2 \df^2\vr
a(\vr, \vb_{01}) a^\dagger(\vr, \vb_{01})
\label{eq:v2}
\end{equation}

\noindent which ensure the conservation of probability: when a dipole
branches into two, the original dipole is lost. For a given dipole,
the branching process and the virtual corrections depend only on its
size, not on its position or on the other dipoles that are present.
The probability of finding a configuration made up of dipoles at
positions and sizes $\{\vr_1, \vc_1; \ldots;\vr_n,\vc_n\}$ after the
evolution through a rapidity $y$ of an initial dipole at $\vr_0$ of
size $\vb = \vb_{01}$ is then:

\begin{equation}
P(\{\vr_1, \vc_1; \ldots;\vr_n,\vc_n\}, \vr_0, \vb, y) =
\langle 0 | a(\vr_1, \vc_1) \ldots a(\vr_n, \vc_n)\e^{y V_R}
a^\dagger (\vr_0, \vb)| 0\rangle
\end{equation}

\noindent where $V_R = V_1 + V_2$. There is a similar expression
for the left-moving onium.

The amplitude $F^{(k)}(\vr,Y)$ for the exchange of $k$ pomerons is
then, analogously to eq.~\ref{eq:f1mine},

\begin{equation}
F^{(k)}(\vr, Y) = \langle 0 | \e^{a_1 + d_1} \frac{(-f)^k}{k!}
\e^{y V_L + (Y-y)V_R} d^{\dagger}(\vr + \vr_0, \vb_L)
a^\dagger(\vr_0,\vb_R)
| 0 \rangle .
\end{equation}

\noindent The term $a_1$ is

\begin{equation}
a_1 = \int \df^2\vr \df^2 \vc \; a(\vr, \vc)
\end{equation}

\noindent with an analogous expression for $d_1$.  The operator $f$
gives the sum of all the dipole-dipole interactions occurring for that
impact parameter:

\begin{equation}
f = \int \df^2\vrc \df^2\vrc'
        \df^2\vc \df^2\vc'
        f(\vrc- \vrc', \vc, \vc')
        d^\dagger(\vrc, \vc) d(\vrc, \vc)
        a^\dagger(\vrc', \vc') a(\vrc', \vc') .
\end{equation}

\noindent Summing together all orders of pomeron exchange, one obtains
an expression which explicitly satisfies the unitarity bound: the
$S$-matrix is

\begin{equation}
S(\vb,\vb',\vr,Y) = 1 + F(\vb,\vb',\vr, Y) =
\langle 0 | \e^{a_1 + d_1} \e^{-f}
\e^{y V_L + (Y-y)V_R} d^{\dagger}(\vr + \vr_0, \vb)
^\dagger(\vr_0,\vb')
| 0 \rangle .
\end{equation}

\noindent Where it would otherwise be unclear which quantity is being
discussed, the unitarised amplitude $F$ will be referred to as
$F^{(unit)}$. This can be cast into a form more suitable for use with a
Monte Carlo simulation. Let $\gamma$ be a particular dipole
configuration for the left moving onium, which contains dipoles of
position and size $(\vr_1,\vc_1) \ldots (\vr_{n_\gamma},
\vc_{n_\gamma})$. For an onium at impact parameter $\vr$, of size
$\vb$, and evolved to a rapidity of $Y$, the probability of finding
such a configuration is defined as $P_\gamma(\vb, \vr, Y)$.  In these
terms, the $S$-matrix becomes:

\begin{equation}
S(Y,\vr,\vb,\vb') = \sum_{\gamma, \gamma'}
        P_{\gamma}(\vr_0, \vb, y)
        P_{\gamma'}(\vr_0 + \vr, \vb',  Y-y)
        \exp(-f_{\gamma, \gamma'}).
\label{eq:mcunit}
\end{equation}

\noindent The definition of $f_{\gamma, \gamma'}$ is:

\begin{equation}
f_{\gamma, \gamma'} = \sum_{i = 1}^{n_{\gamma}}
        \sum_{j = 1}^{n_{\gamma'}}
        f(\vr_{i} - \vr_{j}', \vc_{i}, \vc_{j}')
\label{eq:sumamps}
\end{equation}

\noindent With this form for the interaction, the cross section can be
obtained by randomly producing the configurations for each onium,
and working out $f_{\gamma, \gamma'}$ for each point in impact
parameter. This is repeated with many configurations to obtain the
average $S$-matrix. Since $f_{\gamma, \gamma'}$ is always positive, it
is obvious from eq.~\ref{eq:mcunit} that the unitarity bound is
satisfied.

\subsection{Division of the rapidity interval}
\label{sc:rapdiv}
As already mentioned, for one pomeron exchange, the division of the
rapidity interval between the two onia is arbitrary. The onium-onium
scattering amplitude is independent of $y$ in eq.~(\ref{eq:f1mine}).
This is because $y$ is not a physical parameter. For multi-pomeron
exchange, though $y$ is still not a physical parameter, it does become
a parameter of the approximation, because one is ignoring corrections
to wave functions which have been evolved to $y$ and $(Y-y)$
\cite{Muel94b}: the fractional correction to the forward amplitude
from 2 pomeron exchange is $O(\alps^2 e^{\apm Y})$. For any given part
of the wave function, the leading correction to that part of the wave
function is of the order of $O(\alps^2 e^{\apm y})$ (or $O(\alps^2
e^{\apm (Y-y)})$ for the other wave function). If one divides the
rapidity equally between the two onia, $y = Y/2$, corrections to the
wave functions do not become important, in principle, until twice the
rapidity where corrections appear in scattering. However if one
divides the rapidity interval asymmetrically (e.g. $y \simeq Y$),
corrections to the wave function which have been neglected are of the
same same order as corrections to scattering amplitude and one will
underestimate the total unitarity corrections. This of course ignores
any of the details of the impact parameter distribution of the wave
function, as well as fluctuations which might cause occasional larger
dipole densities, but the basic argument should still be valid.

\section{Some properties of the wave function}
\label{sc:wvfn}
\subsection{Monte Carlo calculation of the wave functions}
The first stage in a Monte Carlo calculation of onium-onium scattering
is to simulate the branching processes which produce the wave
functions of the two onia. For each onium, one starts with an initial
dipole of size $b$ and produces a gluon with a random position whose
distribution is determined by eq.~(\ref{eq:v1}). The rapidity of the
gluon is chosen with an exponentially decaying distribution in
accordance with the virtual corrections of eq.~(\ref{eq:v2}). The
procedure is then repeated with the two new dipoles, and then the
dipoles they produce, etc., until the rapidity of any new gluons would
exceed the rapidity to which one is evolving. The main difficulty in
implementing this procedure is that the integrals in
eqs.~(\ref{eq:v1}) and (\ref{eq:v2}) are divergent for small dipole
sizes. This divergence has to be regulated, and the most appropriate
method for a Monte Carlo simulation is the introduction of a lower
cutoff ($c_{low}$) on the dipole size. In any observable quantity the
divergence cancels out between the real and virtual terms, and the
dependence on the cutoff should disappear as it is taken to zero. For
a finite cutoff, though, any observable will be somewhat modified. The
problem with reducing the cutoff to zero is that the number of dipoles
in the wave function is $\propto 1/c_{low}$. When considering
onium-onium collisions the time taken to calculate an amplitude is the
proportional to the product of the number of dipoles in each onium,
and so is $\propto 1/c_{low}^2$. As a compromise, all evolutions will
be carried out with a lower cutoff of $0.01b$. This will give good
results over most of the rapidity range.  The value of $\alps$ used
will be $0.18$, corresponding to a scale of about $10$GeV.

Before looking at onium-onium scattering, there are certain
properties of the wave function which are worth highlighting.

\subsection{The spatial distribution of dipoles}
An approximate form for the average dipole distribution has been given
in \cite{Muel94b}:

\begin{equation}
n(c,b,r,y) \simeq \frac{2b}{cr^2}
        \log \left(\frac{r^2}{bc}\right)
        \frac{\exp[\apm y - \log^2(r^2/bc) / ky]}
             {(\pi k y)^{3/2}}.
\label{eq:MuelSptl}
\end{equation}

\begin{figure}[htb]
\begin{center}
\epsfig{file=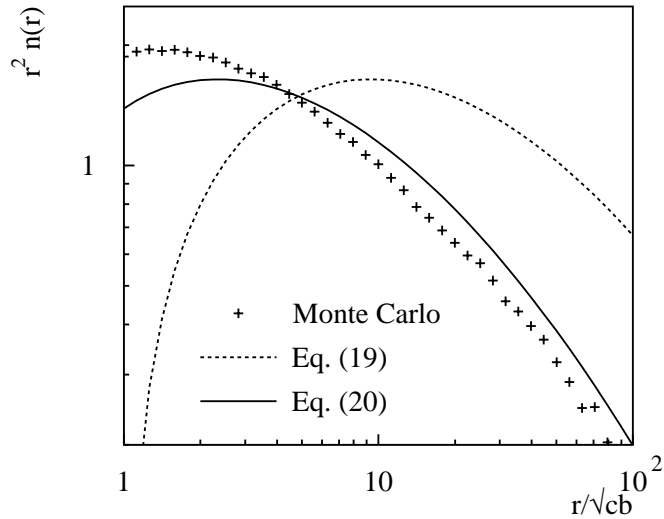, width = 0.6\textwidth}
\caption{The spatial distribution of dipoles in the onium
wave function, for $y = 14$ with $c = b$.}
\label{fig:wf_ceqb}
\end{center}
\end{figure}

\noindent This is the number density of dipoles of size $c$, at a
distance $r$ from the parent dipole of size $b$, after evolution
through a rapidity $y$. The equation is valid for $r \gg b, c$ and
$\log r^2/bc \ll ky$.  One of its main characteristics is that the
maximum of the dipole distribution is at large $r$ (see the dashed
line of figure~\ref{fig:wf_ceqb}). This would imply that scattering is
taking place at values of $r$ where the average dipole distribution
would be quite dilute (though fluctuations would still have to be
taken into account). It would also have the consequence that one might
reach the dangerous infra-red region, where non-perturbative effects
come into play, somewhat sooner than expected.  The Monte Carlo
results (the $+$ symbols on figure~\ref{fig:wf_ceqb}), however, show
that the distribution is far more central.

The source of this discrepancy is a missing portion of one of the
integrals used in the derivation of eq.~(\ref{eq:MuelSptl}). The
details are discussed in Appendix~A, where it is shown that a better
approximation has an extra factor in the logarithms:

\begin{equation}
n(c,b,r,y) \simeq \frac{2b}{cr^2}
        \log \left(\frac{16r^2}{bc}\right)
        \frac{\exp[\apm y - \log^2(16 r^2/bc) / ky]}
             {(\pi k y)^{3/2}}.
\label{eq:mysptl}
\end{equation}

\noindent This modification also applies to the one-pomeron exchange
amplitude, which is why eq.~(\ref{eq:f1sptl}) differs from the result
derived in \cite{Muel94b}.

\subsection{Localised fluctuations}
\label{sc:expflct}
As already discussed, it is not enough to know just the average
distribution of dipoles within the wave function. When calculating
multiple pomeron exchange it is necessary to understand the
fluctuations.  In \cite{Sala95}, it was found that there are very
significant ($P(n)\propto\exp(-\log^2 n)$) fluctuations in the total
number of dipoles of a given size in the onium wave function. The
source of these fluctuations is the occasional production of a large
dipole which cascades down to give many smaller ones. For
multi-pomeron corrections to onium-onium scattering, however, these are
not very important, because the dipoles produced from the large dipole
will be spread over a large region, giving quite a dilute dipole
density. But it was noted that the number of dipoles in a localised
region also has significant fluctuations, this time exponential. As is
discussed in \cite{Muel94b} (and as will be reviewed in section
\ref{sc:toyflct}) an exponential distribution in the dipole density
can cause the series for the amplitudes of $k$-pomeron exchange to be
divergent. Therefore these exponential fluctuations warrant more
study.

A detailed derivation of the exponential nature of the fluctuations is
presented in appendix B. The basis of the derivation is that at very
large $Y$, the $c$ and $b$ dependences factorise, as can be seen from
eq.~(\ref{eq:mysptl}). The $q^{th}$ moment of the distribution of
dipoles of size $c$ inside a region of size $\rho$, from a parent of
size $b$ which is also in the same region, can then be expressed in
the form:

\begin{equation}
n_\rho^{(q)}(c,b,y) = e_q  q! A^q(c/\rho)
                        \left( \frac{b}{\rho} \right)^{\nu_q}
                        \e^{q\apm y},
\end{equation}

\noindent where, for the purposes of the analysis, $A$ can be any
function, though it will actually be proportional to $\rho/c$. All the
moments are defined by the coefficients $e_q$ and the exponents
$\nu_q$ which are approximately independent of $b$. Note that this
form explicitly exhibits KNO scaling \cite{KoNO72} in $c$ and $y$. One
then obtains the following relations, valid for large $q$:

\begin{equation}
\nu_q \simeq 2 - \frac{1}{2q\log 2},
\end{equation}

\noindent and

\begin{equation}
e_q \simeq \frac{1}{4\log 2} \frac{1}{q} \sum_i^{q-i}
        e_i e_{q-i}  \frac{1}{\nu_i + \nu_{q-i} - 2}.
\end{equation}

\noindent Therefore for large $q$, one has $e_q \propto C^q$, where
the exact value of $C$ depends strongly on the behaviour of $e_q$
for low $q$. This gives for the probability of finding $n$ dipoles of
size $c$ within a region of size $\rho$ (in the limit of large $n$):

\begin{equation}
P_n(c, b, \rho, y) \propto \exp\left[
         -\frac{n}{A(c/\rho) C \e^{\apm y}}
        \right].
\end{equation}

\noindent This agrees with the Monte Carlo results to within the
uncertainties, which are due to the accessible values of $y$ being
non-asymptotic and also because the above formulas are valid only for
large $q$, while the value of $C$ retains a dependence on the
behaviour of $e_q$ at smaller $q$.

Note that the shape of the distribution for large $n$ is independent
of the size $b$ of the initial dipole: i.e.\ the shape of the tail of
the distribution is independent of the starting conditions of the
evolution. For example it is possible to produce high density
fluctuations anywhere in impact parameter, as long as a dipole of size
$\simeq \rho$ is produced at that point in impact parameter, early on
in the evolution, so that it has a long range in $y$ to produce many
child dipoles. The only aspect of the tail of the distribution which
does depend on the starting conditions is the normalisation: in the
above example, the probability of producing a dipole of the
appropriate size, at the correct impact parameter, depends on the
position and size of the parent dipole.

\subsection{Consequences of fluctuations: a toy model calculation}
\label{sc:toyflct}
The significance of the exponential density fluctuations is that in
\cite{Muel94b}, Mueller found that exponential fluctuations in a toy
model without transverse dimensions led to divergence of the series
for the amplitude of $k$-pomeron exchange. Consider an onium at a
rapidity $Y/2$, which has a mean number of dipoles $\mu$ ($=\e^{\apm
Y/2}$), with an exponential distribution for the probability of
there being $n$ dipoles:

\begin{equation}
        P_n(Y/2) \simeq \frac{1}{\mu} \e^{-n/\mu}
\end{equation}

\noindent The amplitude $F^{(k)}$ for the interaction of two onia by
exchange of $k$ pomerons is

\begin{equation}
        F^{(k)} = \sum_{m, n} \frac{(-\alps^2 mn)^k}{k!} P_m P_n
\end{equation}

\noindent This is valid for $k \ll m, n$. Substituting the exponential
probability distributions, and approximating the sum by an integral
gives

\begin{equation}
        F^{(k)} \sim k! (\alps\mu)^{2k} (-1)^k
\label{eq:kpomfact}
\end{equation}

\noindent which for $\alps\mu > 1$ is divergent. The same idea applies
in 4 dimensions, as one may see by restricting consideration to the
interaction of dipoles of size $c \sim b$ within regions of size $b$
in each onium (this forces all the interactions to take place at
roughly the same impact parameter, so that they all contribute to
multiple pomeron exchange). These restrictions are appropriate at
small impact parameters where the interaction will be dominated by the
same region in each onium.

\section{Monte Carlo study of onium-onium scattering.}
\label{sc:mcres}

\begin{figure}[p]
\begin{center}
\epsfig{file=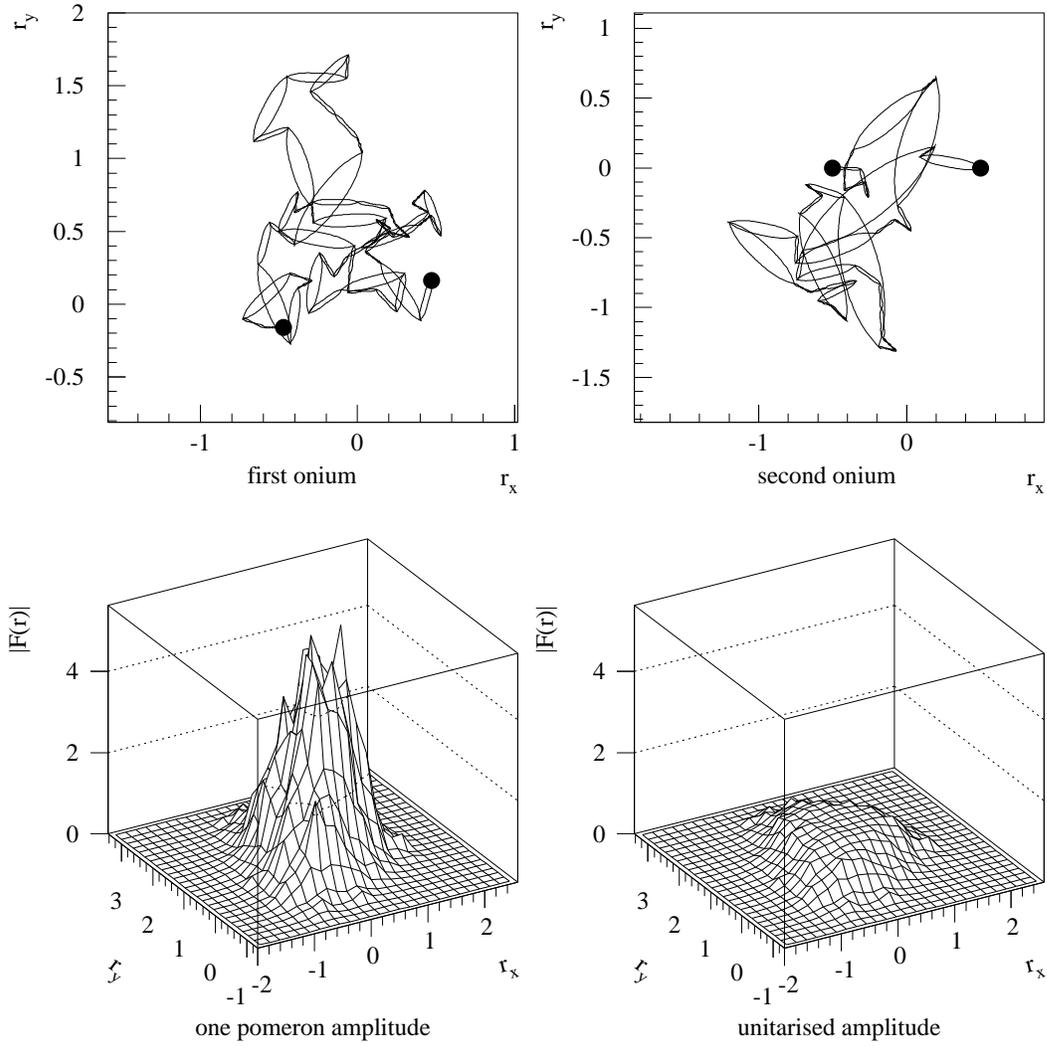, width = \textwidth}
\caption{The dipole structure of two onia, each evolved up to rapidity
$y = 10$ (each `cigar' shape represents a dipole). The black discs
represent the initial quark and anti-quark. The bottom plots
show the amplitude for their interaction as a function of relative
impact parameter.}
\label{fig:evnt_y20}
\end{center}
\end{figure}

Once one has used a Monte Carlo program to generate a pair onium
configurations, as described in section~\ref{sc:wvfn}, one can go on
to look at the interaction between them. This is illustrated in
figure~\ref{fig:evnt_y20}. The top two plots show typical dipole
configurations of two onia. The positions of the initial quark and
antiquark are represented by black discs and each `cigar' shape
represents a colour dipole. The evolution in rapidity results in a
long chain of dipoles stretching between the quark and anti-quark.
The lower left hand plot shows the one-pomeron exchange amplitude as a
function of impact parameter (the position of the
second onium relative to the first). One can see that the largest
interactions occur at an impact parameters where large numbers of
dipoles overlap. At impact parameters where no dipoles overlap, there
is no interaction. The plot is for a high rapidity ($Y = 20$) and most
of the interaction is above 1, and so violates the unitarity limit.
The lower right hand plot shows the unitarised amplitude, which is
quite flat and close to 1: as one would expect, at large rapidities,
the interaction is mostly black.

When determining average amplitudes, two methods have been used. One,
as in figure~\ref{fig:evnt_y20}, involves determining the interaction
at each point of a grid, whose size is chosen to encompass all of the
interaction. This gives an estimate of the total interaction for each
Monte Carlo event. The second approach is to determine the interaction
only along a radial line in impact parameter, from which, after
averaging over many events, one can reconstruct the total amplitude.
Because one is sampling the onium-onium interaction at fewer points in
this second method, the computational time per pair of configurations
is substantially reduced and one can look at a much larger number of
configurations. Full details of the Monte Carlo procedure will be
presented elsewhere\cite{Sala95IP}.

\subsection{1 pomeron, 2 pomeron and unitarised amplitudes}
\label{sc:12uamp}
Before drawing any conclusions from the Monte Carlo calculations, it
is necessary to check that the answers obtained are in good accord
with expectations. All the results which will be given here will be
for the collision of onia of the same size, $b$.

\begin{figure}[htb]
\begin{center}
\epsfig{file=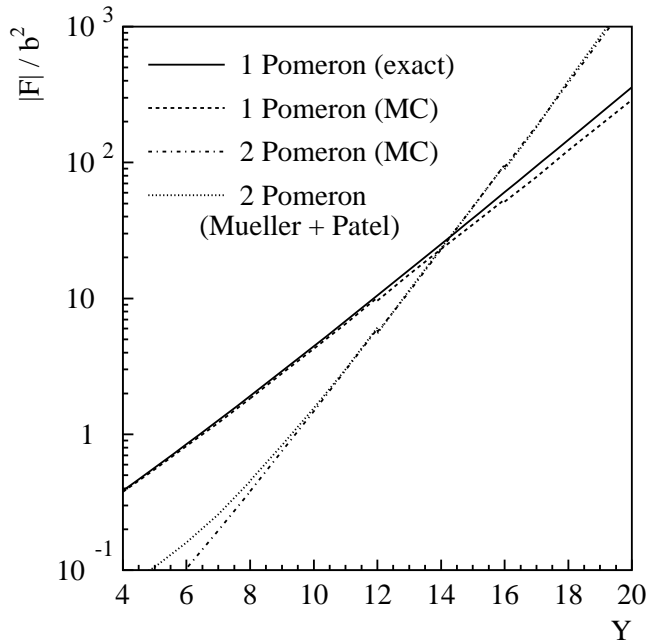, width = 0.6\textwidth}
\caption{The total amplitude for onium-onium scattering as a function of
rapidity. See text for details.}
\label{fig:12p}
\end{center}
\end{figure}

Figure~\ref{fig:12p} shows Monte Carlo results for 1 pomeron exchange,
compared with a result from the numerical solution (as in
\cite{Sala95}) of the differential equation governing the evolution
\cite{MuPa94}. Slight mismatches between Monte Carlo line segments are
visible at $Y=12$ and $Y=16$: the points between $Y=12$ and $16$ all share
the same evolution up to $Y=12$, so that the errors due to
fluctuations in the onium wave function are correlated, making it
easier to see the trends in the evolution. The points from $Y=16$ to
$20$ share their evolutions up to $Y=16$, but they use a
different set of evolutions from the points $Y=12 - 16$, hence the
mismatch at $Y=16$ (and analogously at $Y=12$), which is a measure of
the statistical error. Note that the fluctuations in the wave functions
increase at larger $Y$, while the accessible statistics go down
(because of the larger number of dipoles in the onia), so that the
statistical error increases at large $Y$. This is translated into a
downwards shift of the amplitude, because one usually misses out on
fluctuations with very large amplitudes which would tend to increase
the average amplitude. In addition, at higher rapidities a suppression
of the same order sets in due to the lower cutoff on dipole size.
These small factors aside, the Monte Carlo simulation gives a good
estimate of the amplitude.

The main point of figure~\ref{fig:12p} is to show the relation between
the two-pomeron and the one-pomeron amplitudes. This is of interest
because the rapidity at which the two amplitudes are the same should be
related to the rapidity at which unitarity corrections become
important. The cross over occurs at $Y \simeq 14$, corresponding to an
energy which is well beyond the reach of current accelerators.

In \cite{MuPa94}, Mueller and Patel derived the following relation
for the total two pomeron amplitude:

\begin{equation}
F^{(2)}(Y) = \lambda \left(\frac{[F^{(1)}(Y)]}{\pi k Y}\right)^2.
\label{eq:f2muel}
\end{equation}

\begin{figure}[htb]
\begin{center}
\epsfig{file=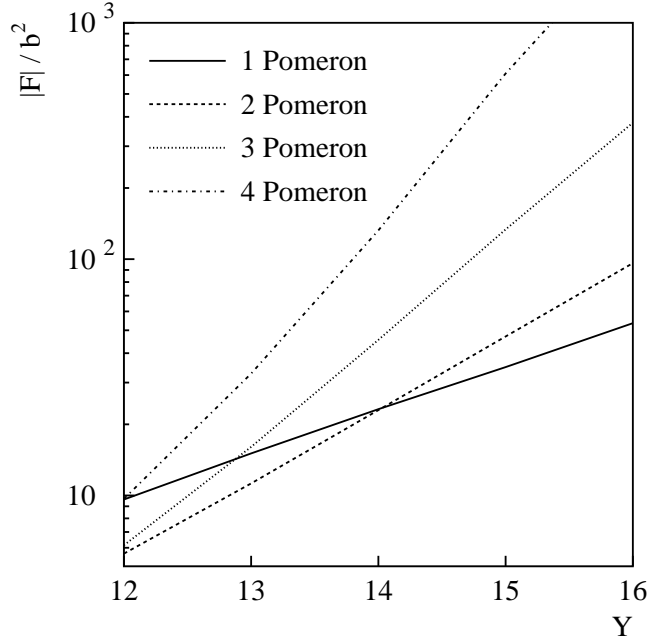, width = 0.6\textwidth}
\caption{Forward, $k$-pomeron amplitude for onium-onium scattering as
 a function of rapidity, as obtained from the Monte Carlo simulation.}
\label{fig:14p}
\end{center}
\end{figure}

\noindent The expression which was given for $\lambda$ was left
unevaluated.  Eq.~(\ref{eq:f2muel}) is plotted in
figure~\ref{fig:12p}, using the Monte Carlo value for $F^{(1)}$, with
a fitted value of $\lambda = 6.9\times 10^2$ (including uncertainties
due to the cutoff, there is an error on this of about $5\%$).
The agreement is very good, except at low rapidities where
non-asymptotic contributions are important.

The results of sections \ref{sc:expflct} and \ref{sc:toyflct}
suggest that, beyond the point where one and two pomeron exchange
cross over, the $k$-pomeron exchange series will be divergent and
hence of no use in calculating the total cross section. This is
clearly visible in figure~\ref{fig:14p} which shows $1$ to $4$-pomeron
exchange for $Y = 12$--$16$. Above $Y = 14$ the series looks
divergent, while at lower rapidities, it looks asymptotic, with $2$
and $3$ pomeron exchange crossing over before $1$ and $2$ pomeron
exchange, and so on and so forth for higher $k$.

One can test the ideas of section~\ref{sc:toyflct} in more detail by
plotting $F^{(k)}(r = 0)/F^{(k-1)}(r = 0)$. In the toy model discussed
there, this would be equal to $k(\alps\mu)^2$, giving a straight line
when plotted against $k$ ($\mu$ was the mean number of dipoles in the
onium without transverse dimensions). The same should apply for the
case with 2 transverse dimensions, except that $(\alps\mu)^2$ will be
replaced by some factor which should be proportional to
$F^{(1)}(r=0)$. The complications due to the transverse dimensions
mean that it is not possible to make any prediction about the constant
of proportionality.

\begin{figure}[htb]
\begin{center}
\epsfig{file=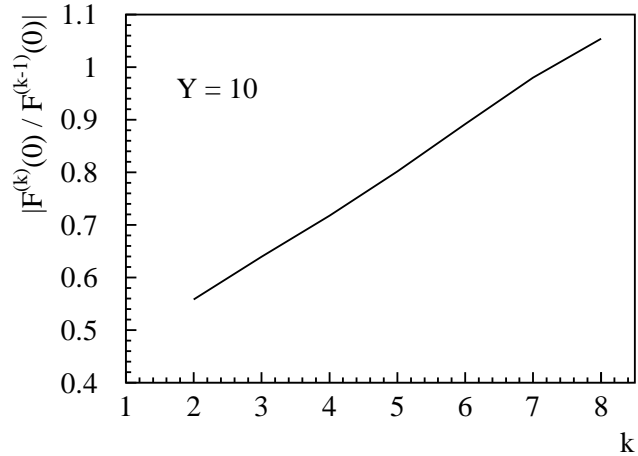, width = 0.6\textwidth}
\caption{Ratio of $k$ to $(k-1)$-pomeron exchange amplitude for impact
  parameter $\vr = 0$.}
\label{fig:cntrl_rts}
\end{center}
\end{figure}

Figure~\ref{fig:cntrl_rts} shows the ratio $F^{(k)}(r = 0)/F^{(k-1)}(r
= 0)$ for $Y = 10$. A low value of $Y$ is chosen to allow the greater
statistics which are needed to sample the tails of the probability
distributions, which are relevant for large $k$. As one can see, the
ratio has a clear linear dependence on $k$. The intercept with the $k$
axis is at non-zero $k$, indicating that the $k$ pomeron amplitude
behaves as $(k + n)!$, rather than $k!$, where $-n\simeq -3$ is the
value of the intercept. This difference means that, after taking into
account the effects of the transverse dimensions, the fluctuations of
the wave functions effectively deviate from exponential at low
multiplicities, which is not unreasonable (the calculation showing the
fluctuations to be exponential was valid only for the tails of the
fluctuations). If one examines plots of the same ratio for other
values of $Y$ (in the range $Y = 8$ to $12$) one finds that the
slope of the straight line is approximately equal to $\beta
F^{(1)}(r)$, where $\beta \simeq 0.2$. From the linear dependence of
the ratio on $k$, and the behaviour of its slope with $Y$, one
concludes therefore that the $k$-pomeron exchange amplitude at $r=0$
is roughly proportional to $(k+n)!  [\beta F^{(1)}(0)]^k$, confirming
that when unitarisation corrections start to become important, the
multiple pomeron series diverges.

\begin{figure}[htb]
\begin{center}
\epsfig{file=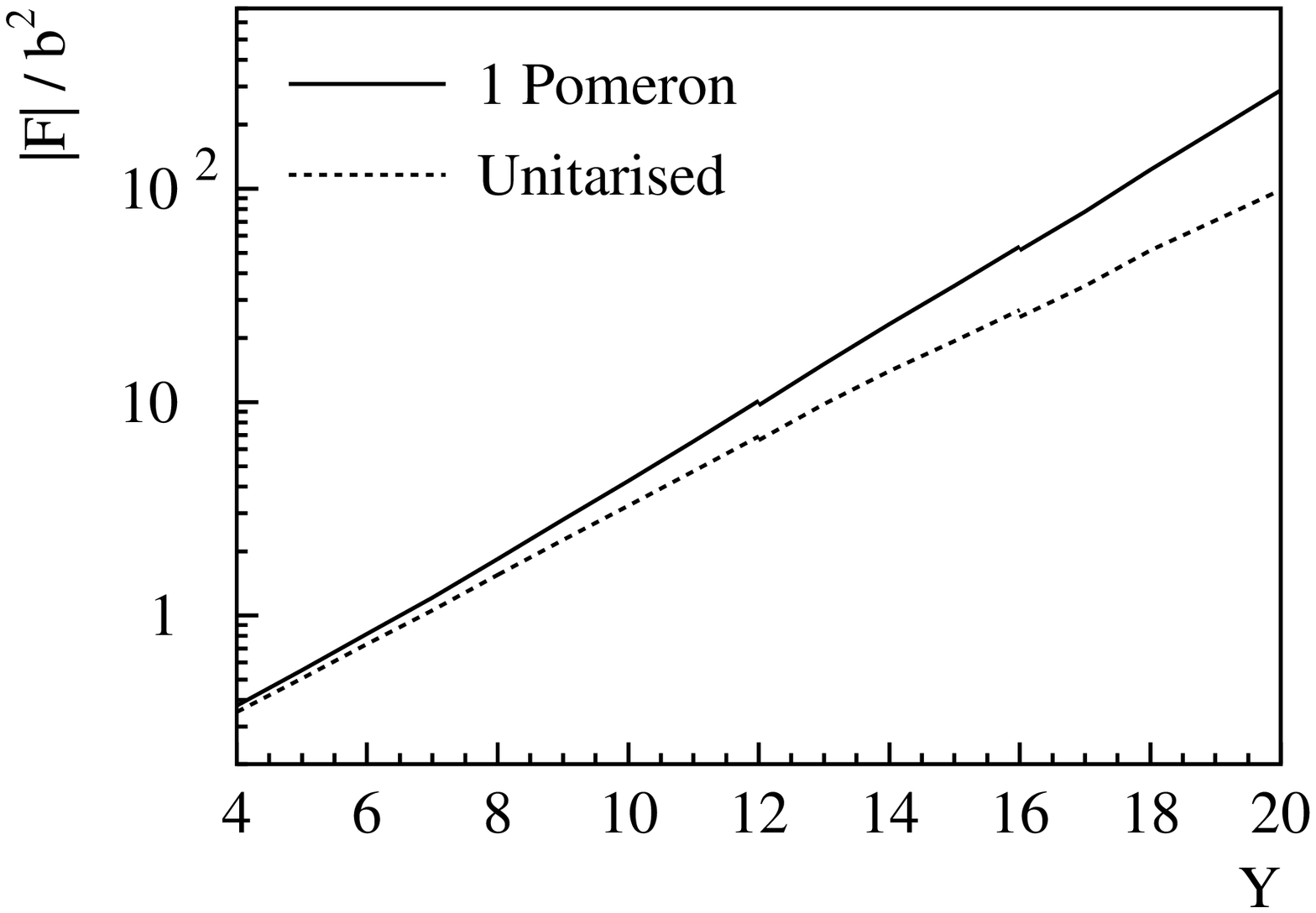, width = 0.6\textwidth}
\caption{Total 1-pomeron and unitarised onium-onium scattering
amplitudes, as obtained from the Monte Carlo simulation.}
\label{fig:unit}
\end{center}
\end{figure}

So at high rapidities, to understand unitarity corrections it is
necessary to use the full unitarised amplitude, resumming multiple
pomeron exchange before summing over configurations of the two onia.
The results of this are shown in figure~\ref{fig:unit}.  Unitarity
corrections set in very gradually: even when the amplitude for two
pomeron exchange is many times larger than the one pomeron amplitude,
the unitarised amplitude carries on growing relatively fast.  At a
rapidity of $Y \simeq 19$, the reduction in the effective power is
only about $25\%$ (see figure~\ref{fig:pwrsvy}). The key to
understanding why unitarity corrections set in so gradually is in the
the profile of the amplitude in impact parameter.

\subsection{The profile of the amplitude: $F(r)$.}
\begin{figure}[htb]
\begin{center}
\epsfig{file=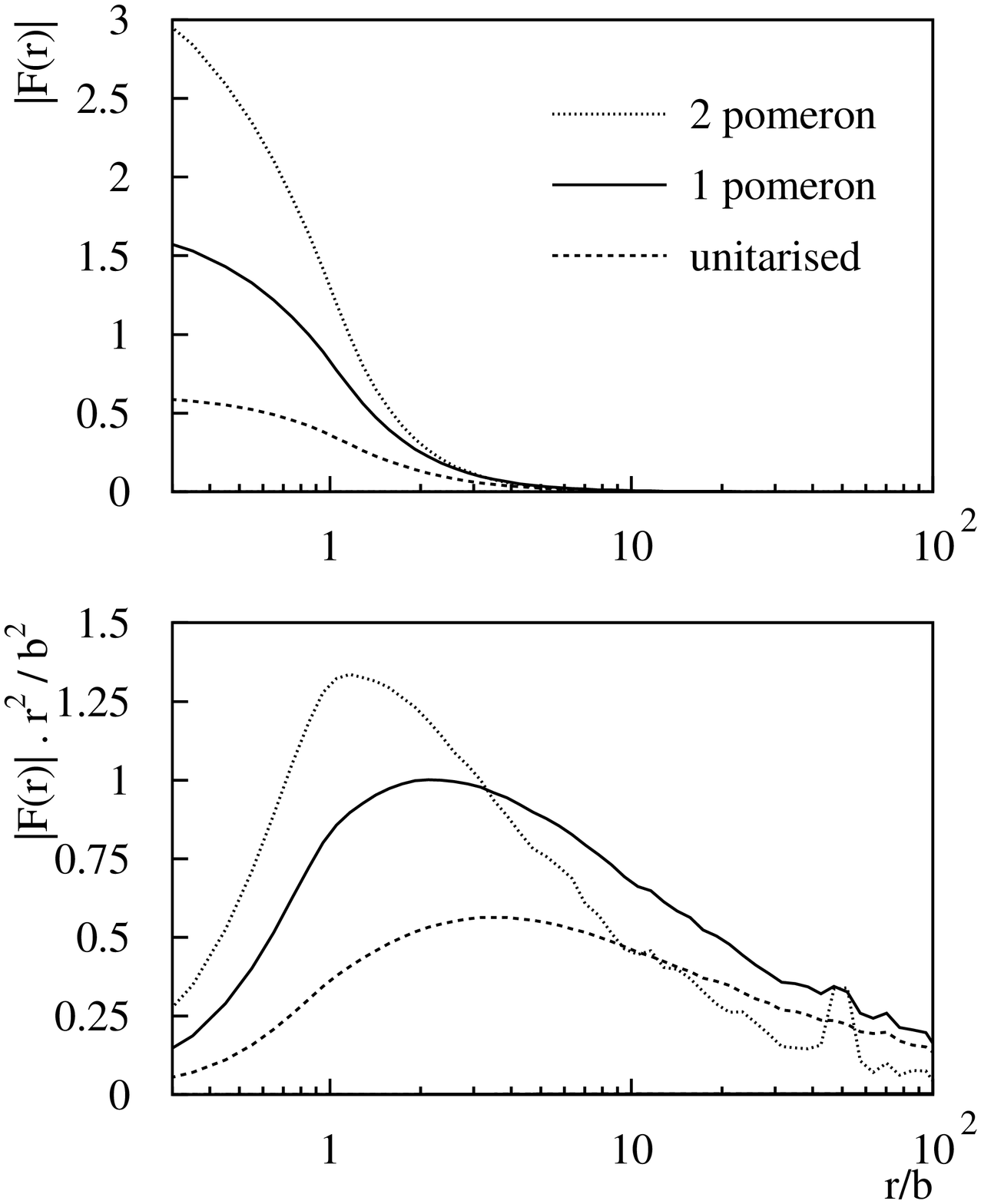, width = 0.6\textwidth}
\caption{The top plot shows the average amplitude as a function of
$r$. In the lower plot, the amplitude is weighted with $r^2$ so that
the area under the curve is proportional to the cross section.
$Y = 14$.}
\label{fig:4prf}
\end{center}
\end{figure}

The profile in impact parameter of the amplitude, $F(r)$,
allows one to determine which region contributes most to the cross
section, and also which region which has the largest unitarity
corrections. The top plot of figure~\ref{fig:4prf} shows $F(r)$
against $r$. As one would expect, the amplitude is largest at small
$r$ and dies off quickly at larger $r$. However, to understand the
contribution to the total amplitude, it is necessary to weight the
amplitude with integration area:

\begin{equation}
        F = \int \df \log r \; 2\pi r^2 F(r).
\end{equation}

\noindent The lower plot of figure~\ref{fig:4prf} shows the
amplitude weighted with $r^2$, so that the total amplitude is
proportional to the area under the curve. The largest contribution to
the total amplitude does not come from the region where $F(r)$ is
largest: the $r^2$ enhancement means that radii of the order of 2 or 3
times the parent size are the most important for the total amplitude,
even though the average amplitude is considerably smaller there than at
$r = 0$.

The two-pomeron amplitude is also shown in figure~\ref{fig:4prf}. With
increasing $r$, it dies off faster than the one pomeron amplitude,
which means that when weighted with $r^2$, it is more central. This is
as suggested by Mueller in \cite{Muel94b}, and the consequence is that
the one-pomeron and two-pomeron amplitudes are dominated by different
regions of impact parameter: the two-pomeron amplitude can become
large without there being large unitarity corrections, because the
two-pomeron contribution (and unitarity corrections) grows mostly near
$r=0$, while the total amplitude comes predominantly from larger
radii. Note that while $F^{(2)}(r)/F^{(1)}(r)$ decreases with $r$, the
ratio $F^{(2)}(r)/[F^{(1)}(r)]^2$ increases quite rapidly with $r$, as
was predicted in \cite{Muel94b} (in fact $F^{(2)}(r)$ decreases even
more slowly with increasing $r$ than was suggested there).

Looking again at figure~\ref{fig:4prf}, one sees that the largest
unitarisation corrections are at also small $r$, so that the
unitarised amplitude, once weighted with $r^2$, has its maximum at
larger $r$ than the one pomeron amplitude. However, the unitarised
amplitude is still considerably suppressed (compared to the one
pomeron amplitude), even at radii where $F^{(1)}(r) \ll 1$. This is
indicative of large fluctuations, or alternatively that eikonalisation
is a poor approximation (and is also related to the strong rise in
the ratio $F^{(2)}(r)/[F^{(1)}(r)]^2$ with increasing $r$). The eikonal
approximation is valid in the limit of no fluctuations: if
$F^{(1)}(\vr)$ is the same for each event then unitarisation can be
done after averaging over events:

\begin{equation}
-F^{unit}(r) \simeq -F^{eik}(r) = 1 - e^{F^{(1)}(r)},
\end{equation}

\begin{figure}[htb]
\begin{center}
\epsfig{file=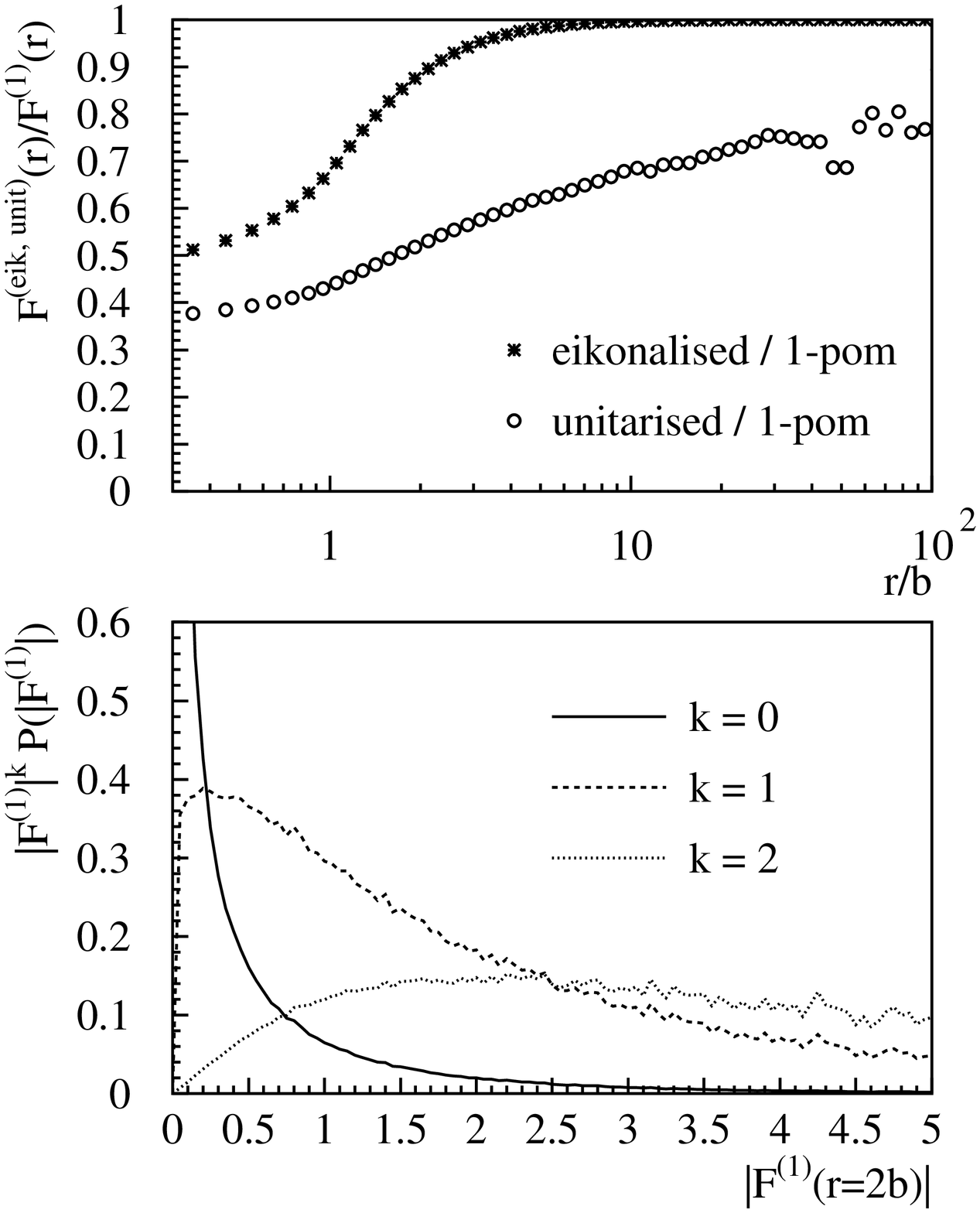, width = 0.6\textwidth}
\caption{Top plot: the ratio of the unitarised and eikonalised
to the one pomeron amplitude, as a function of $r$. Bottom plot: the
probability distribution of $F^{(1)}$ for a fixed impact
parameter, $|\vr| = 2b$. Both plots are for $Y=14$}
\label{fig:eikff}
\end{center}
\end{figure}

\noindent bearing in mind that $F^{(1)}$ is
negative. Figure~\ref{fig:eikff} shows the ratios $F^{unit}/F^{(1)}$
and $F^{eik}/F^{(1)}$. For very small $r$ where the amplitude is quite
black, the eikonalised and unitarised amplitudes are both strongly
reduced, and to a similar extent. They do differ slightly, and this is
the form of failure of eikonalisation found in the toy model
\cite{Muel94b}. However if one examines the radii which contribute
most to the total amplitude ($r\simeq 2b \to 3b$), eikonalisation
would predict only very small corrections, whereas the unitarised
cross section is reduced by a factor of two roughly, compared to the
one pomeron value. This means that the configurations contributing
most to the one-pomeron cross section actually have relatively large
amplitudes. Therefore for the average amplitude to still be low, the
one-pomeron amplitude must be dominated by rare configurations.

The lower plot of figure~\ref{fig:eikff} shows that this is in fact
the case. The solid curve is the probability distribution of the
amplitude $F^{(1)}(\vr)$ for $|\vr|=2b$. The dashed curve shows this
probability distribution weighted by $F^{(1)}(r)$, normalised so that
the total area under it is $1$. The area under a particular region
of the curve is indicative of how much that region contributes to the
average amplitude at $\vr$. The unweighted distribution is very
sharply peaked at $F^{(1)}=0$. Yet the weighted distribution is spread
over a wide range of $F^{(1)}$: 80\% of the amplitude comes from from
only 10\% of the configurations. This is at the value of $r$ which
contributes most to the total amplitude. There is also a contrast
between the curves for 1 pomeron and 2 pomeron (i.e.\ weighted with
$[F^{(1)}]^2(r)$) exchange, however it is less marked: the region
contributing 80\% to the two pomeron amplitude contributes 40\%
to the one pomeron amplitude, though this region comes from only 2.5\%
of configurations, so that the two-pomeron amplitude here is dominated
by somewhat rarer configurations than the one-pomeron amplitude.  The
same phenomenon, of rare configurations dominating one and
multi-pomeron exchange is even more accentuated at larger $r$. It is
also worth pointing out that since the total amplitude comes mostly
from moderate values of $r$, it too is dominated by large
fluctuations: at $Y = 14$, two thirds of the total one-pomeron
amplitude comes from only $10\%$ of events.

The main reason why the amplitude at a given impact parameter is
dominated by such a small fraction of configurations is that for an
impact parameter somewhat away from the onium, dipoles are most likely
to be first produced there later on in the evolution. But if they
appear there only late then they won't have a sufficient rapidity
range to build up in numbers. However if a dipole is produced there
early on (an unlikely occurrence) then it will cascade into many
dipoles and give a large interaction. These rare, (and at larger $Y$,
strongly unitarised) configurations dominate the amplitude, so that
even though the amplitude will on average be small, it can have
significant unitarity corrections. In addition these rare
configurations will not have circular symmetry, so that even when a
state of size $r$ is produced, a point $\vr$ will only have an
interaction if the orientations of the configurations are appropriate,
further increasing the fluctuations in the amplitude at a given point.

What allows the amplitude to continue growing at large rapidities, is
that, for any given impact parameter, the probability of the configurations
with large interactions can carry on increasing until they become
common. But there will always be larger radii at which large
interactions are not yet common, so that the total amplitude can carry
on increasing through an effective increase in the total area of
interaction. This ignores of course the effect of saturation of the
wave function, or confinement effects placing a maximum limit on
the dipole size.

\subsection{Elastic scattering}
\begin{figure}[htb]
\begin{center}
\epsfig{file=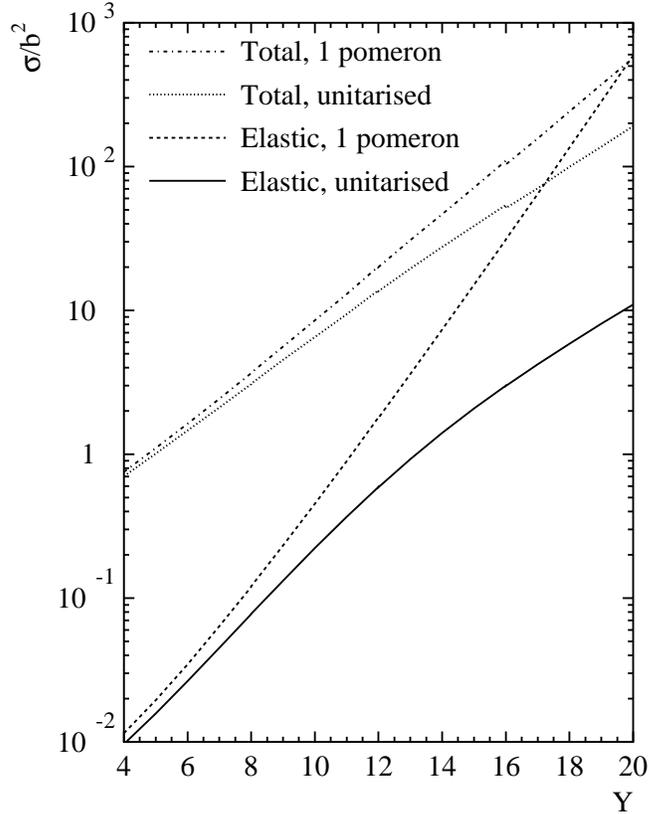, width = 0.6\textwidth}
\caption{The elastic and total cross sections for onium-onium
scattering, as a function of rapidity, showing both the one-pomeron
approximation and the fully unitarised results.}
\label{fig:sigeltot}
\end{center}
\end{figure}

\noindent The unitarity corrections to the total amplitude (and
correspondingly, total cross section)  set in quite slowly,
because the increase in the effective area of interaction allows the
total cross section to carry on rising. However the amplitude is
strongly unitarised at small impact parameters, and concentrating on
this region should make it much easier to see the onset of
unitarity. A crude way of doing this is to look at the elastic
onium-onium cross section. An expression for the differential elastic
cross section was given in eq.~(\ref{eq:elvt}), but it is best to look
initially at the integrated elastic cross section which can be written:

\begin{equation}
\sigma_{el} = \int \df^2 \vr |A(\vr, Y)|^2
\end{equation}

\begin{figure}[htb]
\begin{center}
\epsfig{file=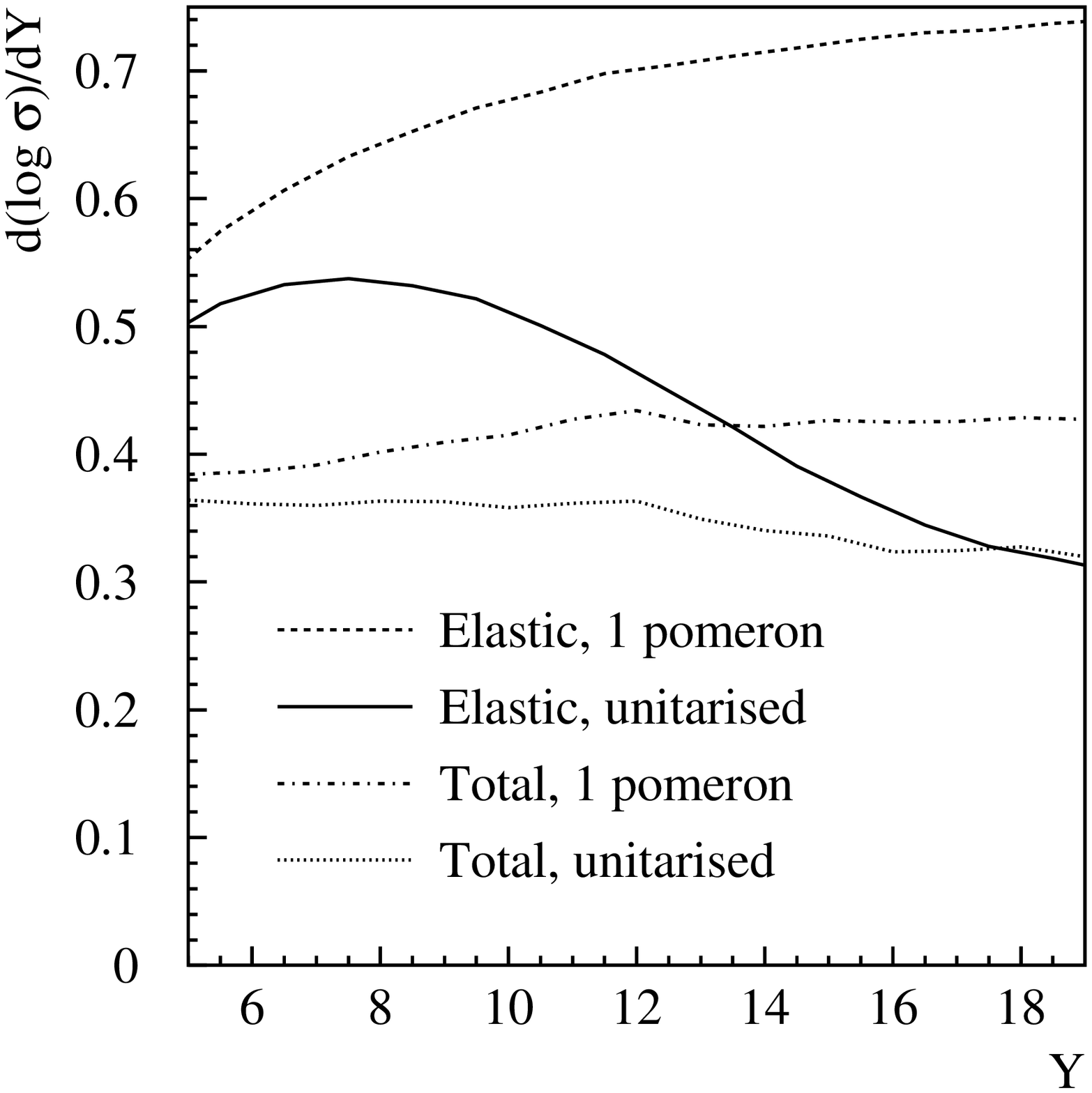, width = 0.6\textwidth}
\caption{The power dependence in rapidity of the total and elastic,
1-pomeron and unitarised cross sections, as a function of rapidity.}
\label{fig:pwrsvy}
\end{center}
\end{figure}

\noindent One should remember that $A(\vr, Y)$ is obtained from
$F(\vb,\vb',\vr,Y)$ by integrating over $\vb$ and $\vb'$ in the
$q\bar{q}$ component of the wave functions of the two onia. However,
for the purpose of understanding the essentials of the behaviour of
the elastic cross section, this complication will be ignored: $|F|$
will be used instead of $|A|$. The leading power behaviour of $F$ is
$1/r^2$, which is why for the total cross section, large impact
parameters can contribute. For the elastic cross section, though, one
is integrating the square of the amplitude, which will fall off as
$1/r^4$. So the integral is dominated by small $r$, where the
unitarisation is strongest. Another aspect of the problem, is that the
elastic cross section can't exceed half the total cross section (the
unitarity condition), but the asymptotic power growth with $Y$ of the
elastic cross section (in the one pomeron approximation) is twice that
of the total cross section. Therefore unitarisation must reduce the
power growth of the elastic cross section by more than a factor of
two.

Figure~\ref{fig:sigeltot} shows the one pomeron and unitarised
calculations for the elastic cross section, together with the
analogous results for the total cross sections, for reference. As
suggested, the unitarisation corrections to the elastic cross section
are much larger than those to the total cross section, and one clearly
sees the rate of increase of the elastic cross section tailing off to
take the same value as that of the total cross section. One also sees
that at large rapidity, the fraction of the total cross section which
comes from elastic scattering is relatively small (about 0.06),
certainly well below the unitarity bound of $1/2$. The reason for this
is, again, that the total cross section is coming from a large region
in impact parameter where the amplitude is small, and so does not
contribute much to the elastic cross section. It is also worth briefly
noting, in view of the discussion on eikonalisation in the previous
section, that the rapidity where the one-pomeron total and elastic
cross sections cross over ($Y\simeq 20$), is the same as the point
where the one and two pomeron contributions to the total cross section
would cross if the eikonal approximation were valid.

In figure~\ref{fig:pwrsvy}, the effective power dependence of the
total and elastic cross sections is plotted against $Y$. Looking first
at the 1 pomeron cross sections, one sees that they vary with $Y$,
tending only relatively slowly to their asymptotic values of 0.47 and
0.94 respectively. This is a result of the logarithmic corrections
which are $1/\sqrt{Y}$ and $1/Y^3$ for the total and elastic cross
sections. The power for the unitarised total cross section deviates
only slowly from the 1-pomeron power, and its variation with $Y$ is
also slow. The unitarised elastic cross section, on the other hand,
grows with a power which differs quite noticeably from the 1-pomeron
elastic cross section, and the variation of the power with $Y$ is also
very different, decreasing rather than increasing from $Y=8$ onwards,
tending towards the same power as the unitarised total cross section
at large rapidity. This significant decrease in the elastic power, if
not masked by other effects, might be a clear signal for the onset of
unitarity corrections.

\begin{figure}[htb]
\begin{center}
\epsfig{file=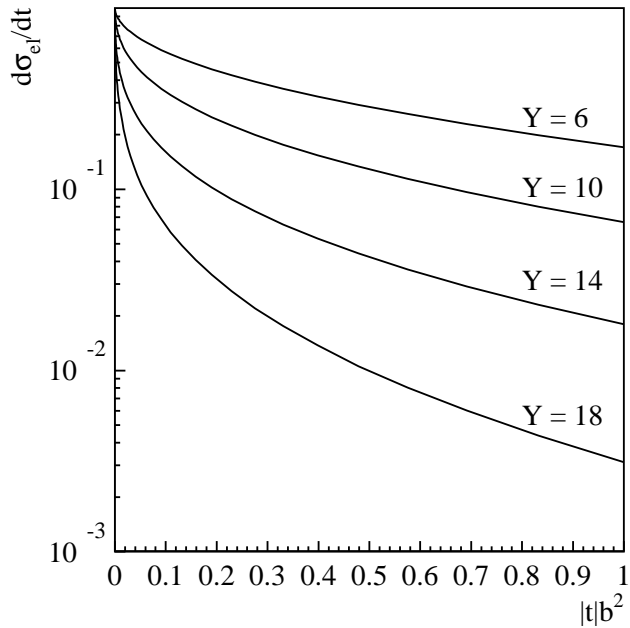, width = 0.6\textwidth}
\caption{The unitarised differential elastic cross section for
different values of $Y$.}
\label{fig:elvt}
\end{center}
\end{figure}

Finally, for completeness, it is worth looking at the differential
elastic cross section, figure~\ref{fig:elvt}, as a function of the
square of the exchanged momentum, $t$. It is very strongly peaked at
$t=0$, because in impact parameter the amplitude dies off so slowly at
large $r$. As $Y$ increases, the region of interaction gets larger,
which causes the elastic peak to become narrower in $t$.

\section{Conclusions}
By performing a Monte Carlo simulation of the dipole formulation of
onium-onium elastic scattering, it has been possible to calculate
single and multi-pomeron scattering amplitudes as a function of
rapidity and impact parameter. The energy where one and two pomeron
total amplitudes are comparable, has been determined to be $Y\simeq
\log s/M^2 \simeq 14$ (with $\alps = 0.18$). However, the multiple
pomeron series is not a very good way of calculating unitarity
corrections. There are two reasons for this. Firstly, the multiple
pomeron series diverges. This is a consequence of the dipole density
in an onium wave function having exponential fluctuations. For multiple
pomeron exchange, the enhancement due to multiple coupling to these
very dense configurations more than compensates for their low
probability. This necessitates the resummation of all numbers of
pomeron exchange before averaging over onium configurations.
Nevertheless, one might have expected that once, say, two pomeron
exchange became larger than one pomeron exchange, the unitarised
amplitude would quickly stop growing.  It doesn't: unitarity
corrections set in very slowly, with the correction to the effective
power of the growth being only of the order of $25\%$, even at a rapidity
of $Y = 19$.

The main reason for this is that the growth of the total multiple
pomeron amplitude is due mostly to an increase in the interaction at
central impact parameters. The unitarised amplitude, on the other hand
effectively carries on growing through an increase in the area of
interaction, even when it has already reached the unitarity bound at a
given point in impact parameter. This though is still a
simplification. One finds that the impact parameters contributing most
to the total amplitude have low average amplitudes, yet significant
unitarity corrections. In fact, the average amplitude there is
dominated by rare configurations with large amplitudes and large
unitarity corrections: the growth of the total amplitude comes through
an increase in the probability of these rare configurations.

Since the unitarisation is largest at small impact parameters, it is
useful to find a process where small impact parameters matter
more. Elastic scattering has this property because it depends on the
square of the average amplitude so regions with the largest amplitudes
(and strongest unitarisation) contribute most. One finds that the
unitarisation effects start to become important as early as $Y = 8$:
the one-pomeron approximation to $\df \log\sigma_{el}/\df t$ carries
on increasing with $Y$, while the unitarised power starts to decrease.

There are a number of difficulties when trying to relate these results
to experimentally feasible measurements. Ideally one would want to
examine a process such as that suggested by Mueller and Navelet
\cite{MuNa87}, where the BFKL evolution occurs between two objects
with similar, large, transverse scales. However, the evidence
currently for BFKL type behaviour is in results from HERA, both in the
measurement of the rise of the structure function $F_2$ at small~$x$
\cite{H1F295,ZEF295} (analogous to a total cross section) and more
recently from the rise (roughly twice as fast) of the elastic
electro-production of $\rho^0$ and $J/\psi$ and elastic
photo-production of $J/\psi$
\cite{ZErhoe95,ZEpsie95,H195rhoe,H195psip,H195psie} (analogous to
elastic scattering). Complications arise in trying to understand the
unitarity corrections in these types of processes, partly because the
evolution is taking place over a range of scales, but also because of
the non-perturbative effects associated with the proton scale. In
particular, these would prevent the large size fluctuations, which are
so important for the continued growth of the BFKL total cross section
beyond the point where unitarisation becomes significant.

\section*{Acknowledgements}
I am very grateful to A.H.~Mueller and B.R.~Webber for many helpful
conversations and for suggesting this work, as well as to M.~Diehl,
F.~Hautmann, R.~Peschanski and W.-K.~Tang for useful discussions.

\newpage
\appendix
\section*{Appendix A: The mean spatial distribution of dipoles.}
The discrepancy between the Monte Carlo results for the dipole
distribution and the analytical prediction from Mueller \cite{Muel94b}
is related to the details of one of the integrals involved in its
evaluation. In determining eq.~(\ref{eq:MuelSptl}) one goes through
the following stage:

\begin{equation}
n(c,b,r,y) = \int \frac{\df\nu}{2\pi} H_{\nu}(\vc,\vb,\vr)
        \frac{b}{c}
        \exp\left[\frac{2\alps N_C}{\pi}\chi(1 + 2i\nu) y\right]
\end{equation}

\noindent where
\begin{equation}
H_\nu(\vc,\vb,\vr) = \frac{4\nu^2}{\pi^2} b^{-2i\nu} c^{2i\nu}
 \int \frac{\df^2 \vrc}
 {[|\vrc - \vb/2| |\vrc + \vb/2|]^{1-2i\nu}
  [|\vrc + \vr - \vc/2| |\vrc + \vr + \vc/2|]^{1+2i\nu}}
\label{eq:hnuint}
\end{equation}

\noindent The BFKL characteristic function, $\chi(\gamma)$, is

\begin{equation}
\chi(\gamma) = \psi(1) - \frac{1}{2}\psi(1-\gamma/2)
			- \frac{1}{2}\psi(\gamma/2),
\label{eq:bfklchi}
\end{equation}

\noindent where $\psi$ is the digamma function. First, cut out the
regions $R < Pb$ and $|\vrc + \vr| < Pc$, where $P$ is an arbitrary
constant satisfying $P \gg 1$ and $4\nu P \ll 1$. The integral without
these regions gives:

\begin{equation}
\frac{2i\nu}{\pi r^2} \left[
        \left(\frac{r^2}{P^2 bc}\right)^{-2i\nu} -
        \left(\frac{r^2}{P^2 bc}\right)^{2i\nu} \right].
\label{eq:hbigr}
\end{equation}

\noindent One can see, by taking the limit $\nu \to 0$, that there
should be no terms of order $\nu$ within the square bracket, because
the integral is similar to the one evaluated in \cite{Muel94a} for the
virtual corrections to the dipole kernel. The result obtained in
\cite{Muel94b} is equivalent to taking $P = 1$ and neglecting other
contributions. Now consider one of the regions that was cut out. It
can be integrated as follows:

\begin{eqnarray}
& &     \frac{4\nu^2}{\pi^2}
        \frac{b^{-2i\nu}c^{2i\nu}}{r^{2 + 4i\nu}}
        \int^{R<Pb} \frac{R\df R \df\theta}
        {(R\sqrt{b^2 - 2bR\cos\theta + R^2})^{1-2i\nu}} \\
&\simeq &       \frac{4\nu^2}{\pi^2}
        \frac{(P^2bc)^{2i\nu}}{r^{2 + 4i\nu}}
        \int^{R<Pb} \frac{\df R \df\theta}
        {\sqrt{b^2 - 2bR\cos\theta + R^2}}.
\end{eqnarray}

\noindent The approximation giving the second line is valid because
of the condition $\nu P \ll 1$, so that in the range of $R$
important for the integration, the power $\nu$ has little effect on
the value of the integrand. The angular integral can be performed to
give a complete elliptic integral of the first kind,
$[4\mathbf{K}(R/b)/b]$ for $R<b$ and $[4\mathbf{K}(b/R)/R]$ for $R>b$
\cite{Book:GR}. Performing the $R$ integral then gives

\begin{equation}
\frac{8\nu^2}{\pi r^2}
        \left(\frac{r^2}{P^2bc}\right)^{-2i\nu}
        [\log 4P + O(\nu)]
\simeq \frac{2i\nu}{\pi r^2}
        \left(\frac{r^2}{P^2bc}\right)^{-2i\nu}
        \left[ (16P^2)^{-2i\nu} -1+ O(\nu^2)\right].
\end{equation}

\noindent Adding this, and a similar term for the dipole of size $c$,
to eq.~(\ref{eq:hbigr}) gives

\begin{equation}
H_\nu(\vc,\vb,\vr) \simeq \frac{2i\nu}{\pi r^2} \left[
        \left(\frac{16r^2}{bc}\right)^{-2i\nu} -
        \left(\frac{16r^2}{bc}\right)^{2i\nu} \right].
\end{equation}

\noindent Following through the calculation, gives for the dipole
density:

\begin{equation}
n(c,b,r,y) = \frac{2b}{cr^2}
        \log \left(\frac{16 r^2}{bc}\right)
        \frac{\exp[\apm y - \log^2(16 r^2/bc) / ky]}
             {(\pi k y)^{3/2}}.
\end{equation}

\noindent This is much more central and agrees well with the Monte
Carlo distributions.

\section*{Appendix B: Fluctuations in the localised dipole number}

The approach used, will be to study the equations for the moments of
the dipole distribution within a region of radius $\rho$. The first
step is to note that branching sequences that include dipoles of sizes
larger than $\rho$ will not contribute to high density fluctuations:
the number dipoles of size $\rho$ coming from a dipole of size $R \gg
\rho$ will be at most enhanced by a factor proportional to $R/\rho$
(just the usual factor in front of the BFKL saddle point solution).
However dipoles will be spread out over a region of size $R^2$, so
that the density of dipoles will be suppressed by a factor $\rho/R$.
Therefore in calculations one can leave out all dipoles larger than
$\rho$. One can also neglect the transverse positions of the dipoles,
since, if a dipole starts inside a region of size $\rho$, and the
branching involves no dipoles larger than $\rho$, most of the child
dipoles will still be inside the region.

Define $n_\rho^{(q)}(c,b,y)$ to be the $q^{th}$ moment of the number
of dipoles of size $c$ in a region of radius $\rho$ originating from a
parent of size $b$ after an evolution through rapidity $y$. (The
$q^{th}$ moment being $\langle n(n-1)\ldots(n-q+1)\rangle$). It will
then approximately satisfy the following equation:

\begin{equation}
\frac{\df n^{(q)}_\rho(c,b,y)}{\df y} = I_\rho^{(q)} +
        \frac{\alps N_C}{2\pi^2}
        \int^\rho \frac{b^2 \df^2\vb_2}{b_{02}^2 b_{12}^2}
        \left[ n^{(q)}_\rho(c,b_{02},y) + n^{(q)}_\rho(c,b_{12},y) -
                n^{(q)}_\rho(c,b_{01},y)
        \right].
\label{eq:momscut}
\end{equation}

\noindent The inhomogeneous term, $I_\rho^{(q)}$, is

\begin{equation}
        I_\rho^{(q)} = \frac{\alps N_C}{2\pi^2}
        \int^\rho \frac{b^2 \df^2\vb_2}{b_{02}^2 b_{12}^2}
        \sum_{i=1}^{q-1} C_i^q
        n^{(i)}_\rho(c,b_{02},y)n^{(q-i)}_\rho(c,b_{12},y)
\end{equation}

\noindent This is obtained from the generating functional equation in
\cite{Muel94a}, with an upper cutoff placed on the dipole size in the
$\vb_2$ integration.

The first approximation to make relies on the large $Y$ limit. At
large $Y$, with $c$ and $b$ not too different, one has for the
total average number of dipoles:

\begin{equation}
        n(c,b,Y) \propto \frac{b}{c} \e^{\apm Y}
\end{equation}

\noindent The important point is that the $b$ dependence (which shows
a power behaviour) factors out from the $c$ and $Y$ dependences. This
will apply to the moments of the localised dipole distribution as well.
It will further be useful to make the assumption (reasonably justifiable)
that the $b$ dependences of the moments are also power behaviours. One
can then parameterise each moment by two constants $e_q$ and $\nu_q$,
which should depend only weakly on $b$:

\begin{equation}
n_\rho^{(q)}(c,b,Y) = e_q  q! A^q(c/\rho)
                        \left( \frac{b}{\rho} \right)^{\nu_q}
                        \e^{q\apm Y}.
\label{eq:naseq}
\end{equation}

\noindent It is not necessary to know the exact $c$ dependence, which
is why it is left as $A(c)$. With this parameterisation,
eq.~(\ref{eq:momscut}) can be written as:

\begin{equation}
        q \apm \left( \frac{b}{\rho} \right)^{\nu_q} e_q =
        \frac{\alps N_C}{\pi} \left[
        k_{\nu_q} \left(\frac{b}{\rho}\right)^{\nu_q} e_q +
        \left(\frac{b}{\rho}\right)^2
        \sum_{i = 1}^{q-1} e_i e_{q-i} l_{\nu_i\nu_{q-i}}
        \right].
\label{eq:eq}
\end{equation}

\noindent The following integrals have been defined:

\begin{equation}
k_\nu = \frac{1}{2\pi} \left(\frac{b}{\rho}\right)^{-\nu}
        \int^\rho \frac{b^2 \df^2 b_2}{b_{02}^2 b_{12}^2}
        \left[
        \left(\frac{b_{02}}{\rho}\right)^{\nu} +
        \left(\frac{b_{12}}{\rho}\right)^{\nu} -
        \left(\frac{b}{\rho}\right)^{\nu}
        \right],
\end{equation}

\noindent and

\begin{equation}
l_{\nu \nu'} = \frac{1}{2\pi} \left(\frac{b}{\rho}\right)^{-2}
        \int^\rho \frac{b^2 \df^2 b_2}{b_{02}^2 b_{12}^2}
        \left[
        \left(\frac{b_{02}}{\rho}\right)^{\nu}
        \left(\frac{b_{12}}{\rho}\right)^{\nu'}
        \right].
\end{equation}

\noindent For $1 \le \nu < 2$ (as should be the case for all $q$),
$k_\nu$ and $l_{\nu\nu'}$ will depend only weakly on $b/r$ (except for
$l_{11}$ which has a logarithmic dependence on $b/r$ --- this is
discussed later).

The main difference between the two terms is that the $l_{\nu \nu'}$
integral is dominated by large $b_2\sim \rho$, while the $k_\nu$
integral, because of the condition given above on $\nu$, the integral
is dominated by small $b_2\sim b$. This is the origin of the different
power behaviours from the two integrals.

To solve eq.~(\ref{eq:eq}) one uses the fact that $\nu_q < 2$,
therefore for large $\rho/b$ the inhomogeneous term drops out, leaving:

\begin{equation}
q\apm \simeq \frac{\alps N_C}{\pi} k_{\nu_q}.
\end{equation}

\noindent Through the $\nu$ dependence in $k_\nu$, this fixes
$\nu_q$. For moderately large $q$, $k_\nu$ can be approximated as

\begin{equation}
        k_\nu \simeq \frac{2}{2-\nu}
\end{equation}

\noindent giving

\begin{equation}
\nu_q = 2 - \frac{1}{2q\log 2}
\end{equation}

\noindent Note that this is just the solution to the equation
$\chi(\nu) = q\chi(1)$, in the limit $\nu \to 2$, with $\chi$
defined as in eq.~(\ref{eq:bfklchi}). The other region
which can be studied to help solve eq.~(\ref{eq:eq}) is that of $b$
close to $\rho$. For the case of $\nu$ close to $2$, i.e.\ for large
$q$, at large $b$, the integral for $k_\nu$ is much smaller than its
asymptotic value (it only reaches its asymptotic value for $\rho/b
\sim \exp[1/(2-\nu)]$), while that for $l_{\nu\nu'}$ should be much
closer to its asymptotic value (the integrand is approximately
proportional to $b_2$). Therefore the $k_\nu$ term can be neglected,
giving

\begin{equation}
e_q \simeq \frac{1}{4\log 2} \frac{1}{q} \sum_i^{q-i}
        e_i e_{q-i} l_{\nu_i\nu_{q-i}}
\label{eq:eqiter}
\end{equation}

\noindent A reasonable approximation for $l_{\nu_1\nu_2}$ is:

\begin{equation}
l_{\nu\nu'} \simeq \frac{1}{\nu + \nu' - 2}.
\label{eq:lnunu}
\end{equation}

\noindent By iterating eq.~(\ref{eq:eqiter}), one can determine all
the moments of the fluctuations. Its form is such that for large $q$,
$e_q$ should be behave as follows:

\begin{equation}
        e_q \simeq B C^q
\end{equation}

\noindent Because of the complicated behaviour of the coefficients of
the iteration for low values of $q$, it is not possible to provide an
analytical approximation for $B$ and $C$, though they can be
determined numerically. This then gives an exponential distribution
for the fluctuations in dipole number in a restricted region,
$\e^{-n/\mu}$ where

\begin{equation}
\mu(c/r) = A(c/r) C e^{\apm Y}
\end{equation}

\noindent It is difficult to compare this directly with the results
from the Monte Carlo simulation, because $A(c/r)$ contains various
logarithmic factors which are difficult to estimate accurately.
However the mean number of dipoles of a given size in a certain region
is something that can be obtained from the Monte Carlo distribution
and so $A(c/\rho)$ can be factored out:

\begin{equation}
  \frac{\mu(c/\rho)}{n_\rho(c,\rho,Y)} \simeq C \simeq \lim_{q \to
    \infty} \frac{e_q}{e_{q-1}}.
\label{eq:eqeqm1}
\end{equation}

\noindent In evaluating $C$ from the recursion relations there are some
complications because the first relation contains $l_{11}$ which
varies logarithmically with $b/\rho$ (this is the reason why the
expression given in eq.~(\ref{eq:lnunu}) diverges). Given that the
values of $e_q$ are determined by the region of $b$ close to $\rho$,
it is then natural to use a value of $l_{11}$ corresponding also that
region of $b$. For $b = 0.5\rho$, $l_{11} = 1.9$ (determined
numerically), while for $b = 0.25\rho$, $l_{11} = 2.7$. This range
will be used when calculating the $e_q$ to gauge the uncertainty in
the calculation.

Another uncertainty arises because at the rapidities available to the
Monte Carlo simulation, the logarithmic prefactors affect the
effective power growth of the distribution. To gauge this effect, one
can replace $\apm$ with a measured effective power and look at the
difference that this makes to $C$. The final range of values obtained
for $C$ is from $0.34$ to $0.75$.

\begin{figure}[htb]
\begin{center}
\epsfig{file=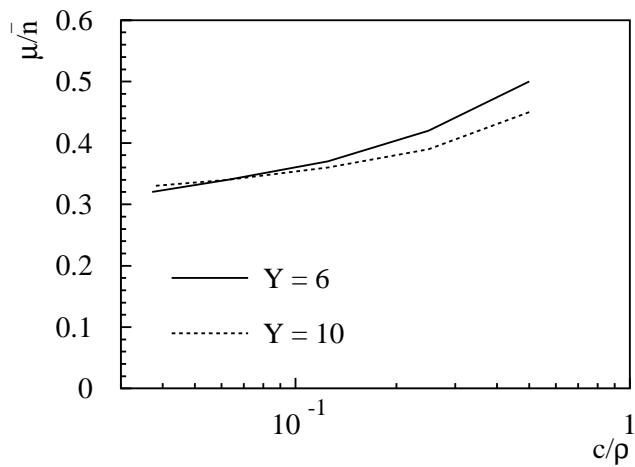, width = 0.6\textwidth}
\caption{The parameter $C$ of eq.~(\ref{eq:eqeqm1}), as determined by
Monte Carlo simulation for two values of $y$.}
\label{fig:muplot}
\end{center}
\end{figure}

Figure~\ref{fig:muplot} shows the value of
$C_{MC}=\mu(c/\rho)/[2n_\rho(c,\rho/2,Y)]$ determined from the Monte
Carlo simulation for two values of $Y$ and a range of $c/\rho$. The
values for $C$ from the iteration of eq.~(\ref{eq:eqiter}) are
consistent with the Monte Carlo results. However the Monte Carlo
results do depend slightly on $Y$ and on $c/\rho$. For the variation
with $Y$ at $c \sim \rho$, $C_{MC}$ decreases as $Y$ increases. This
could be because at larger $Y$, the effective power is larger. A
larger effective power leads to a reduced value for $e_q/e_{q-1}$
(this can be seen by making some simple analytic approximations, and
is also obtained in the numerical iterations). A second effect is
that at smaller $c/\rho$, the presence, non-asymptotically, of factors
of the form $\exp[-\log^2(c/\rho)/kY]$ leads to an increase in the
effective power with decreasing $c$, giving a decreasing $C$. At
larger $Y$ this effect would be smaller, so, at larger $Y$, $C$ should
decrease more slowly with decreasing $c$. Both of these observations
are qualitatively in accord with the Monte Carlo results.


\end{document}